\documentclass[aps,pre,showpacs,superscriptaddress]{revtex4}
\usepackage{times,xspace}
\usepackage{latexsym,amsbsy,amssymb,amsmath,bm}
\usepackage{graphicx,amsmath,amsfonts}
\usepackage{epsfig}
\begin{document}
\title {Role of range of interactions in a model of diffusing charged particles}
\author{Krzysztof Pawlikowski}
\email{kriss@ift.uni.wroc.pl}
\author{Andrzej P\c{e}kalski}
\email{apekal@ift.uni.wroc.pl}
\affiliation{Institute of Theoretical Physics, University of Wroc{\l}aw,\\
pl. M. Borna 9, 50-204 Wroc{\l}aw, Poland}

\begin{abstract}
 Role of range of interactions in a model of charged particles diffusing on a two-dimensional lattice is studied. We investigate, via Monte Carlo simulations, three models. In the first one interactions are restricted to nearest neighbors, in the second one we add (weaker) interactions to second nearest neighbors. In the third model interactions are extended as far as the half the size of the system. The strength of these interactions falls off with the distance.
We show that for static properties, such as asymptotic spatial patterns or the order parameter, the results of the first and third models are very close. Dynamic characteristics, like mean square displacement, average cluster size and average energy as functions of time show larger differences among the three models. Origins of these differences are pointed out and discussed,
\end{abstract}
\maketitle
\section{Introduction}
Motion of particles and pattern formation on planar surfaces is a much studied subject \cite{gomer,gouyet}. Understanding basic mechanisms of such processes was always important, as it describes motion of adatoms or ions on surfaces. It became recently even more so with the growing interest in nanostructures \cite{mirek}. Diffusion of particles on a substrate is a cooperative phenomenon resulting from interactions among many particles, particles with a substrate and from thermal excitations, Theoretical papers devoted to these problems are many. Some of them are based on analytic approach \cite{gouyet,bunde,kehr,kutner}, others used Monte Carlo simulations \cite{sadiq,uebing,he,mertelj}. The range of processes studied includes pattern formation \cite{he,mertelj,tata} or dynamics \cite{gouyet,sadiq,uebing}. In some of these papers the studied systems was composed of one type of particles, in others there were two types, $A$ and $B$ \cite{gouyet,sadiq,he}. Sometimes also vacant sites were allowed. In general, interactions of diffusing particles with the substrate atoms were neglected and the role of the substrate was reduced to creating an unchanging periodic potential acting on the diffusing particles. Forces between the particles were modelled in many ways. Simple nearest-neighbor (NN) interactions \cite{uebing}, next nearest-neighbors (NNN) ones \cite{sadiq,uebing} and even farther \cite{he,mertelj}. The interactions could be of one type (repulsive or attractive) or they could be of both types, depending on the distance between the interacting particles \cite{uebing}.

In two recent papers \cite{my1,my2} we have introduced a simple model of two types of particles moving on a two-dimensional square lattice. The particles are identical, except for having opposite charge. There is the same number of positively and negatively charged particles, hence the net charge is always zero and the dynamics is of the Kawasaki type. A particle is free to move to an empty NN site according to the Metropolis algorithm. Restricting interactions to NN and weaker NNN ones, we have shown how the resulting patterns and dynamic parameters, like diffusion coefficient, mean square displacement, average cluster size depend on the range of interactions (to NN or to NNN), coverage and the temperature.

In this paper we investigate, using Monte Carlo simulations, a problem which is quite important for modeling diffusion on surfaces and which, to the best of out knowledge, has not yet been investigated, namely how important is the restriction of the interactions to NN and NNN. It is well known \cite{melzer} that if the local charge is screened, the range of the potential between interacting particles could be reduced. We shall compare here spatial patterns, order parameters, mean square displacement, tracer diffusion coefficient, average cluster size and average energy for three versions of the basic model introduced in \cite{my1}. In the first version, which we shall henceforth call $H_1$, the interactions are restriced to NN, in the second ($H_2$) one (weaker) interactions to NNN are added. In the third version, called $H_3$, each particle interacts with all particles which are not farther from it than half the size of the system. Therefore, with periodic boundary conditions used by us, each particle interacts with all others, although the strength of such interactions falls off with the distance between the particles, contrary to mean-field models, where the interactions to all particles have the same strength. Quite often restricting the range of interactions of charged particles, acting with Coulomb forces, is criticized. It seems therefore useful to check how different would be the results coming from interactions reduced to NN, then to NNN, as compared with the long range ones.
\section{Model}
A system of positively ($A$) and negatively ($B$) charged particles located at sites of a square lattice of dimensions $L \times L$ is considered. The lattice has periodic boundary conditions and odd number of nodes. The total number of particles of both types divided by the number of sites defines the coverage, which in the following will be equal 0.2. A lattice site could be either empty, or occupied by one $A$ or $B$ particle. Hamiltonians for the three models described above have the form
\begin{equation}
 \mathcal{H}_1\,=\, J_1 \sum_{i=1}^N \sum_{j=1}^{\delta_1} n_i\,n_j\quad ,
\end{equation}
where $N$ is the total number of particles, $J_1$ is a positive constant and $n_i = 0, \,\pm 1$, depending whether the site $i$ is empty, occupied by an $A$ or $B$ particle. $\delta_1$ means summation over first neighbors. When the interactions are extended to NNN the Hamiltonian takes the form
\begin{equation}
 \mathcal{H}_2 \,=\,
 \sum_{i=1}^N  n_i\left( J_1 \sum_{j=1}^{\delta_1}  n_j + J_2\sum_{j=1}^{\delta_2} n_j\right) \, ,
\end{equation}
where $J_2 = J_1/\sqrt{2}$ and $\delta_2$ means summation over NNNs. Finally, when the interactions are up to any particle within the $(L-1)/2$ range, we have
\begin{equation}
\mathcal{H}_3\,=\, \sum_{i\neq  j=1}^N J_{ij}\,n_i\,n_j  .
\end{equation}
Now the sum  extends to neighbors not farther than $(L-1)/2$ and the interaction constant is divided by the distance between the sites $i$ and $j$. The three models belong to the class of stochastic lattice gas models with additional degree of freedom (charge). In one Monte Carlo step (MCS), which is our time unit, each particle could move to an empty NN site according to the standard Metropolis algorithm \cite{landau}. Constant temperature in the system is maintained through a contact with a heat reservoir.

We shall investigate the following quantities: the order parameter, asymptotic, i.e. at the end of simulations, spatial patterns, mean square displacement (msd), average energy and average cluster size as functions of time. From the mean square displacement $\langle r^2(t) \rangle$ the tracer diffusion coefficient $D^*$ is calculated as \cite{gomer}
\begin{equation}
\label{d}
D^*\,=\,\frac{1}{4Nt} \sum_{i=1}^N \langle \left|\vec{r}_i(t) - \vec{r}_i(0)\right|^2 \rangle \, ,
\end{equation}
where $\vec{r}_i(t)$ and $\vec{r}_i(0)$
are the positions of the particle $i$ at time $t$ and time $0$, respectively. Another quantity recorded  is the order parameter $\varphi$  introduced in Ref.~\cite{my1} and generalized to interactions up to $m$ neighbors as
\begin{equation}
\label{f1}
\varphi_m\,=\, \frac{m}{8N} \sum_{i=1}^N \left| \sum_{j=1}^{\delta_1} n_i n_{i+j} -  \sum_{j=1}^{\delta_2} n_i n_{i+j}\right |\, .
\end{equation}
This order parameter is zero in high temperatures (no stable pattern) and is non-zero in low temperatures. Time dependence of the average cluster size yields useful information about the emerging spatial structure of the system and its stability. Following Landau and Binder \cite{landau} a cluster is an agglomeration of minimum 10 particles which have at least one particle as a NN or NNN. Average energy is the sum all interactions in the system divided by the twice the number of particles.

As we have explained before \cite{my2}. we have studied a system of rather modest size, namely 51 $\times$ 51. In many cases to see interesting results we had to run simulations to quite long times, up to $3\times 10^7$ MCS. In large systems processes are going more slowly, hence to observe such effects one would have to run simulations for at least  two orders of magnitude longer times, which  would make it out of our computing powers. We have checked before, Ref. \cite{my2}, that for low coveraged reported in this paper, there are no differences between systems with $L$ = 51 and $L$ = 101. Most of the results showing time dependence of average energy etc. come from single simulation runs. The reason is that characteristics of these curves are, in general, shifted in time in different runs. Averaging will result in smearing out of these effects. We have verified that the presented curves are generic and not just particular effects. All results presented below are for one, rather low, coverage $c$ = 0.2. Discussion of the effects of higher coverages for the models $H_1$ and $H_2$ has been given in Ref. \cite{my1,my2}.
Data for quantities like average energy, cluster size or msd are reported for three sets of temperatures -- low (below the critical temperatures $T_c$), medium (a little above $T_c$) and high (much above $T_c$). The Arrhenius and order parameter plots are given for all interesting temperatures.
\section{Results}
\begin{figure}
\begin{center}
\includegraphics[scale=0.6]{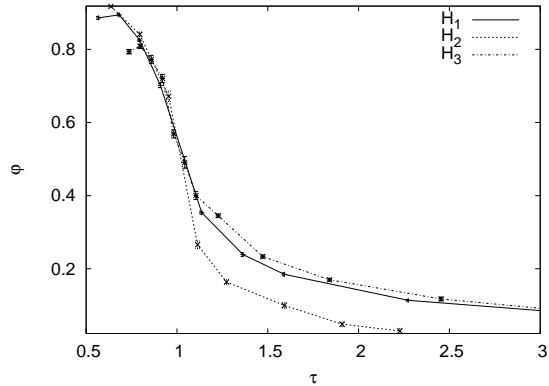}
\end{center}
\caption{Order parameter versus scaled temperature for three models - interactions to NN, up to NNN and the long range ones. Scatter of the data is marked by error bars.}
\label{fi}
\end{figure}

We begin with the order parameter $\varphi$. Quite often, especially in investigating critical phenomena, relevant parameter is not just the temperature, but rather its deviation from the critical value. Since in this paper we compare results of models differing by interactions an appropriate parameter will be the reduced (scaled) temperature, $\tau = T/T_c$, i.e. temperature divided by its critical value. From the dependence of $\varphi$ on $T$ we have estimated $T_c$ as the inflection point of the $\varphi(T)$ curve. The plot of $\varphi(\tau)$ is shown in Figure \ref{fi}.

\begin{figure}
\begin{center}
\includegraphics[scale=0.2]{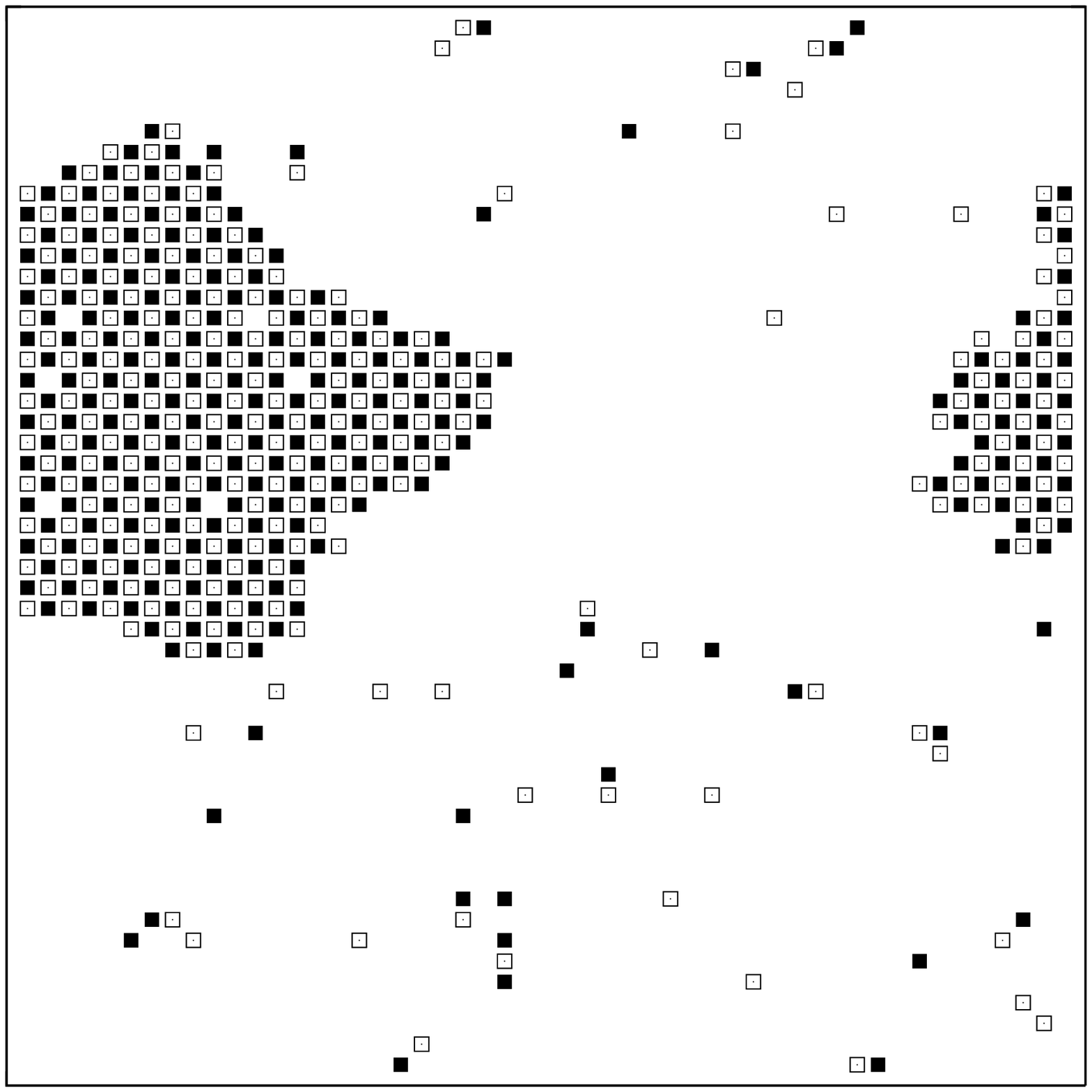}\hspace*{3ex}
\includegraphics[scale=0.2]{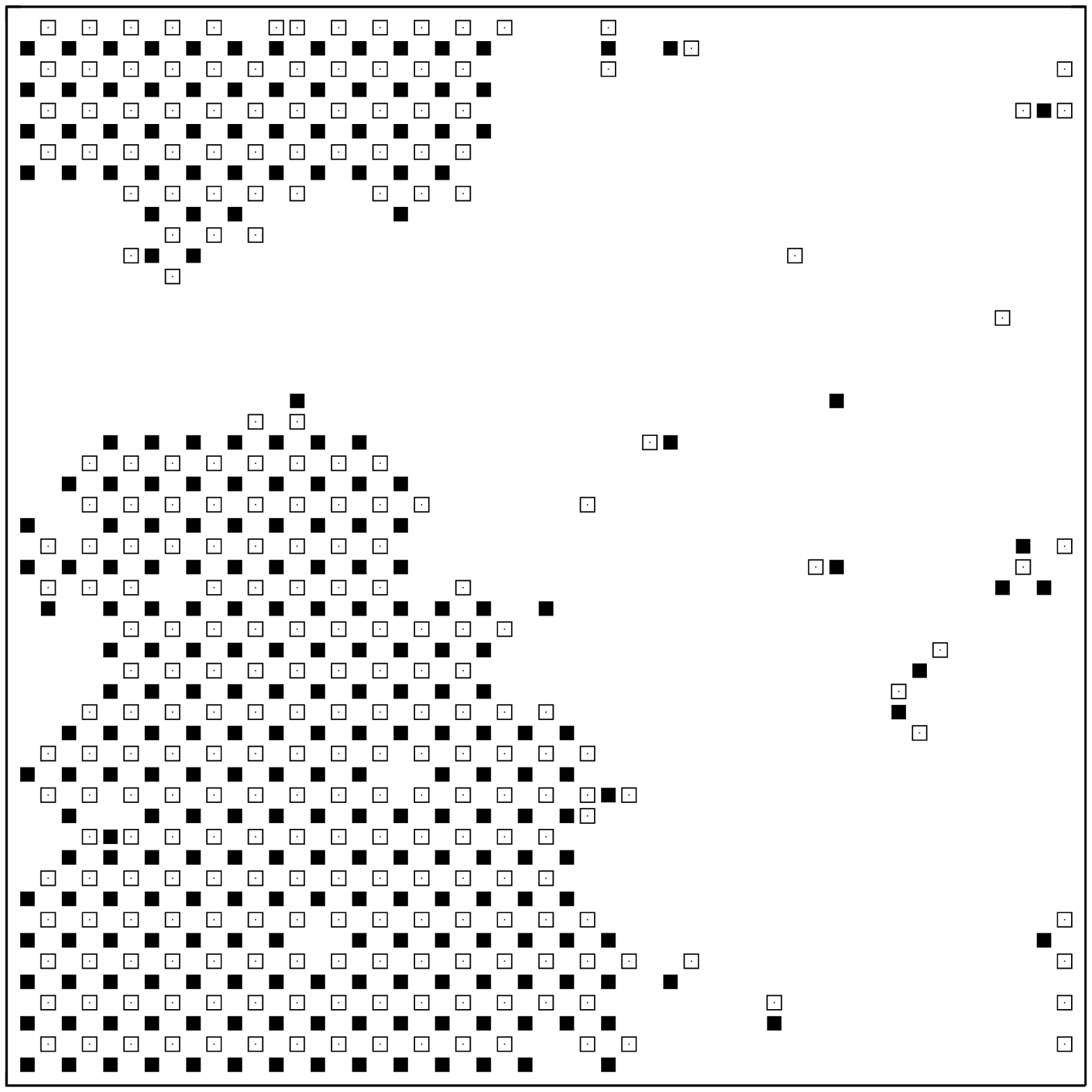}\\[3ex]
\includegraphics[scale=0.2]{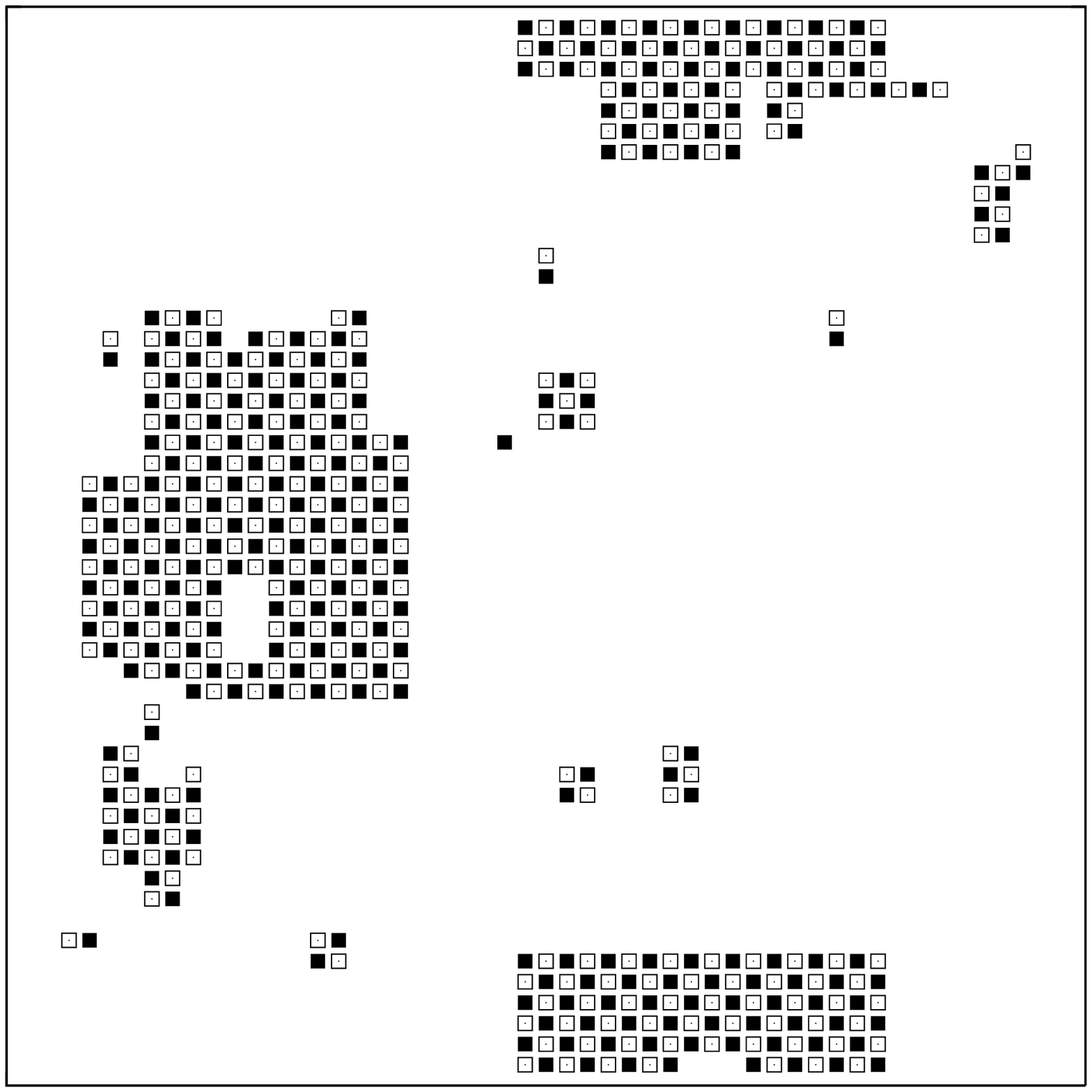}\hspace*{3ex}
\includegraphics[scale=0.2]{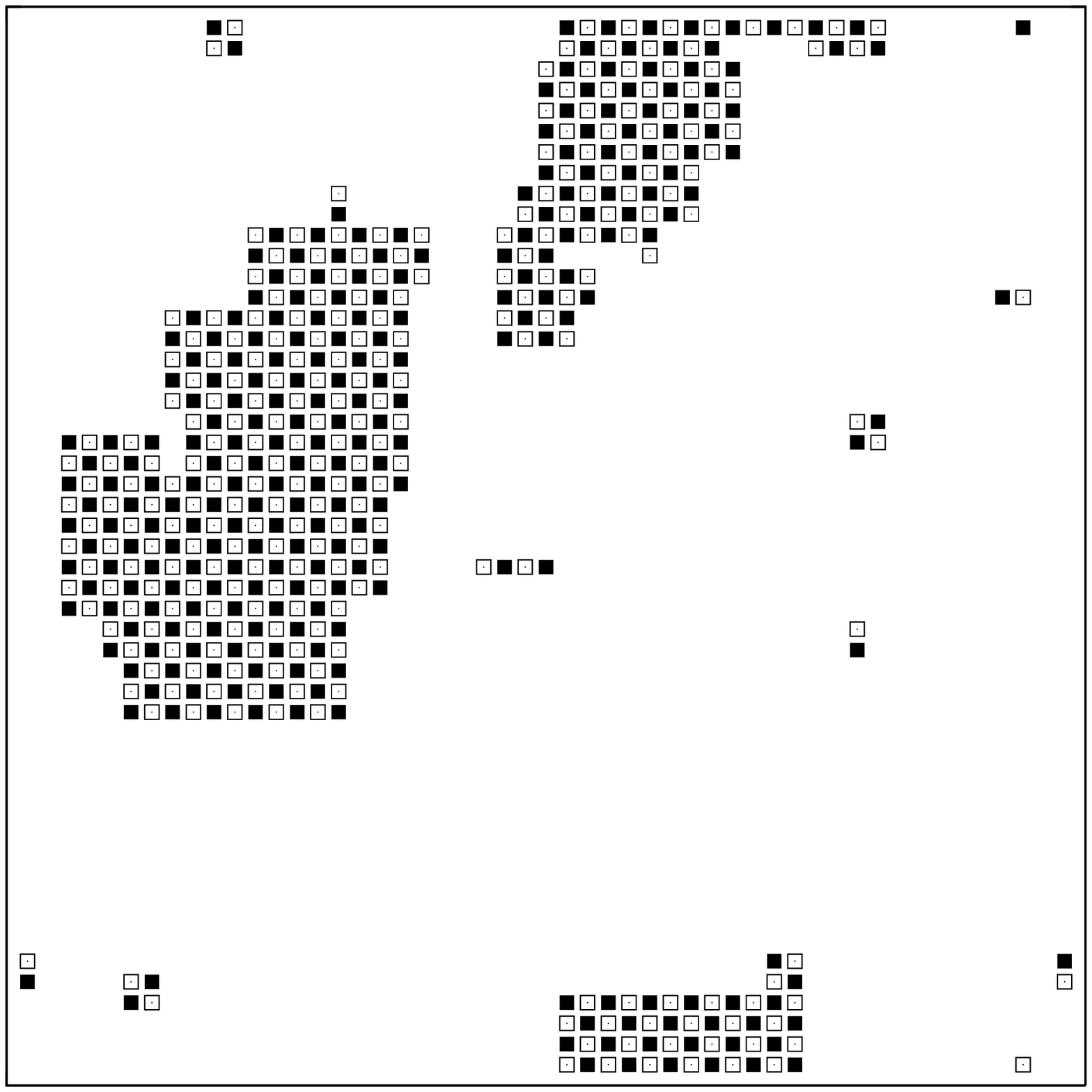}
\end{center}
\caption{Spatial structures obtained at low temperatures after $4\times 10^6$ MCS for the three models: $H_1$ (left upper panel), $H_2$ - (right upper panel), $H_3$ (left bottom panel) and $H_3$ after $3\times 10^7$ MCS (right bottom panel).}
\label{fig2}
\end{figure}

\begin{figure}
\begin{center}
\includegraphics[scale=0.4]{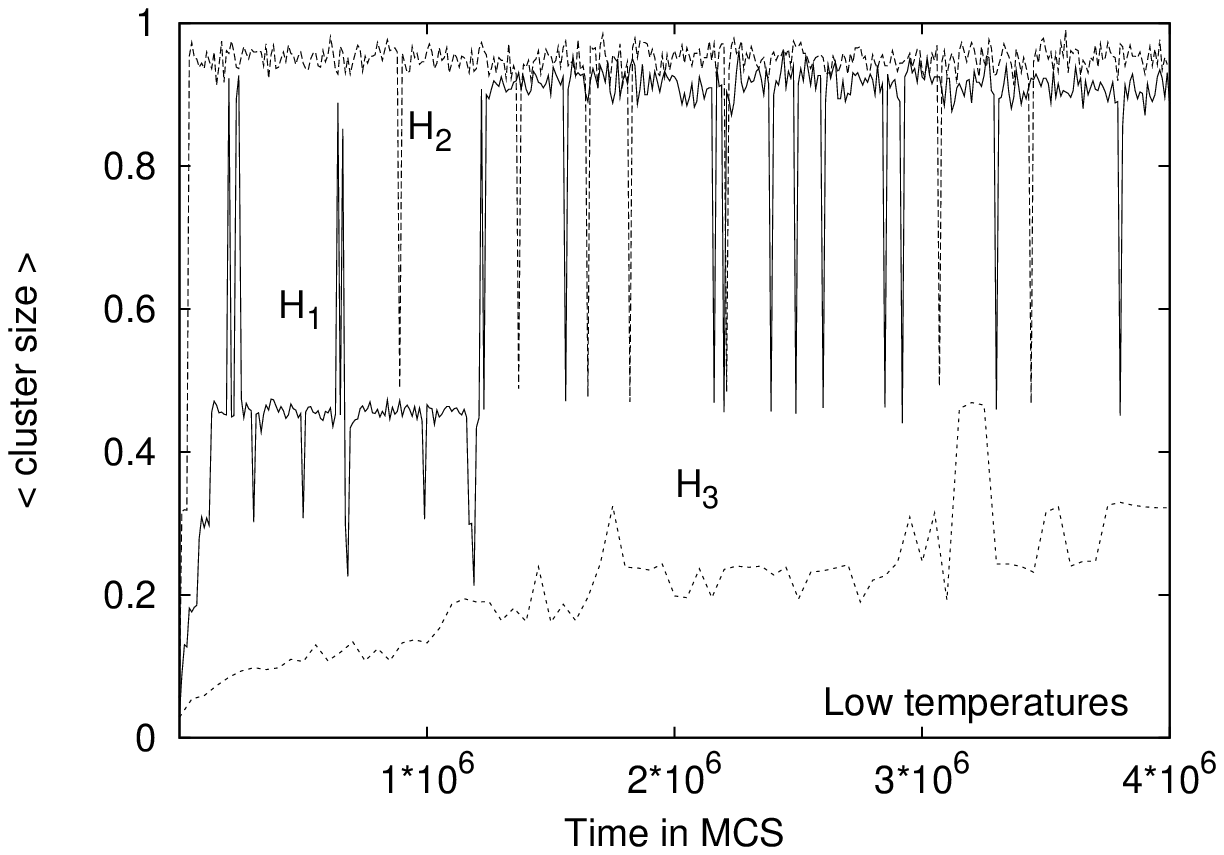}
\includegraphics[scale=0.4]{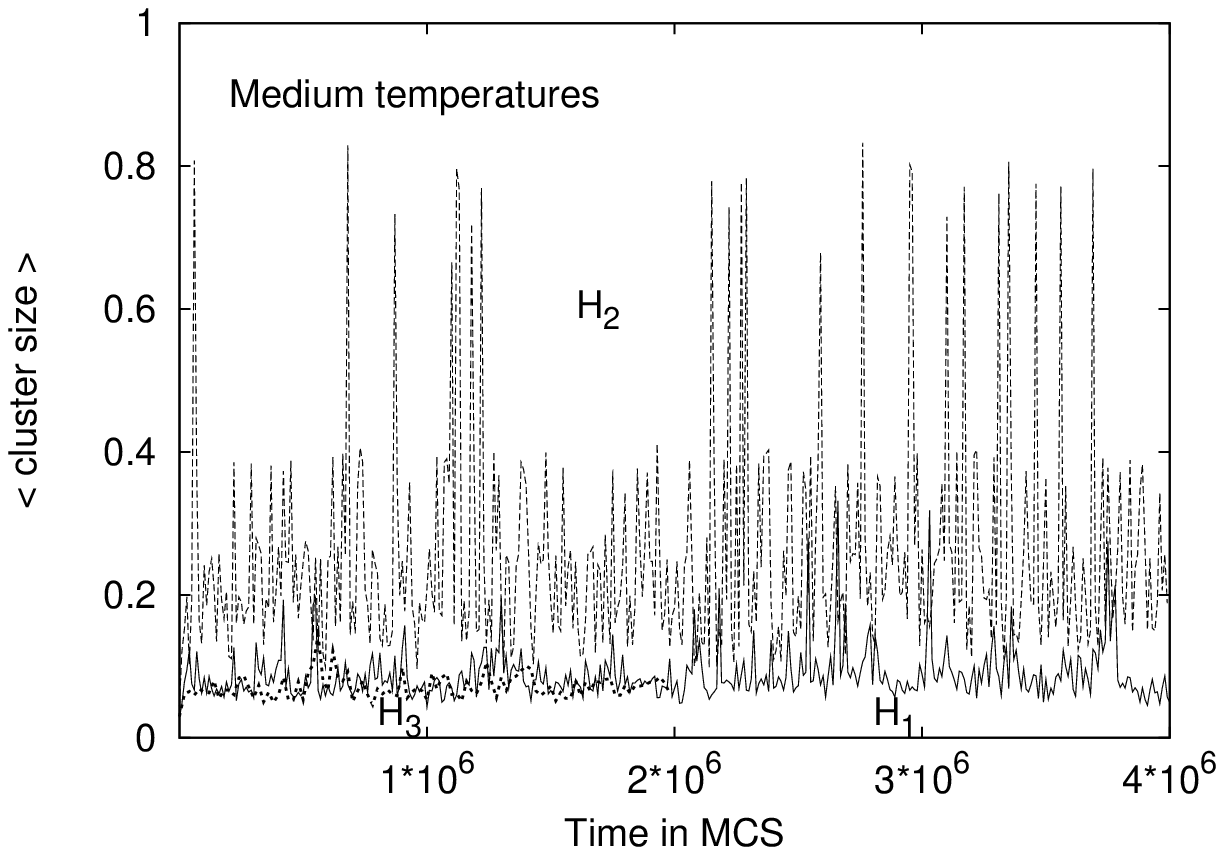}\\
\includegraphics[scale=0.4]{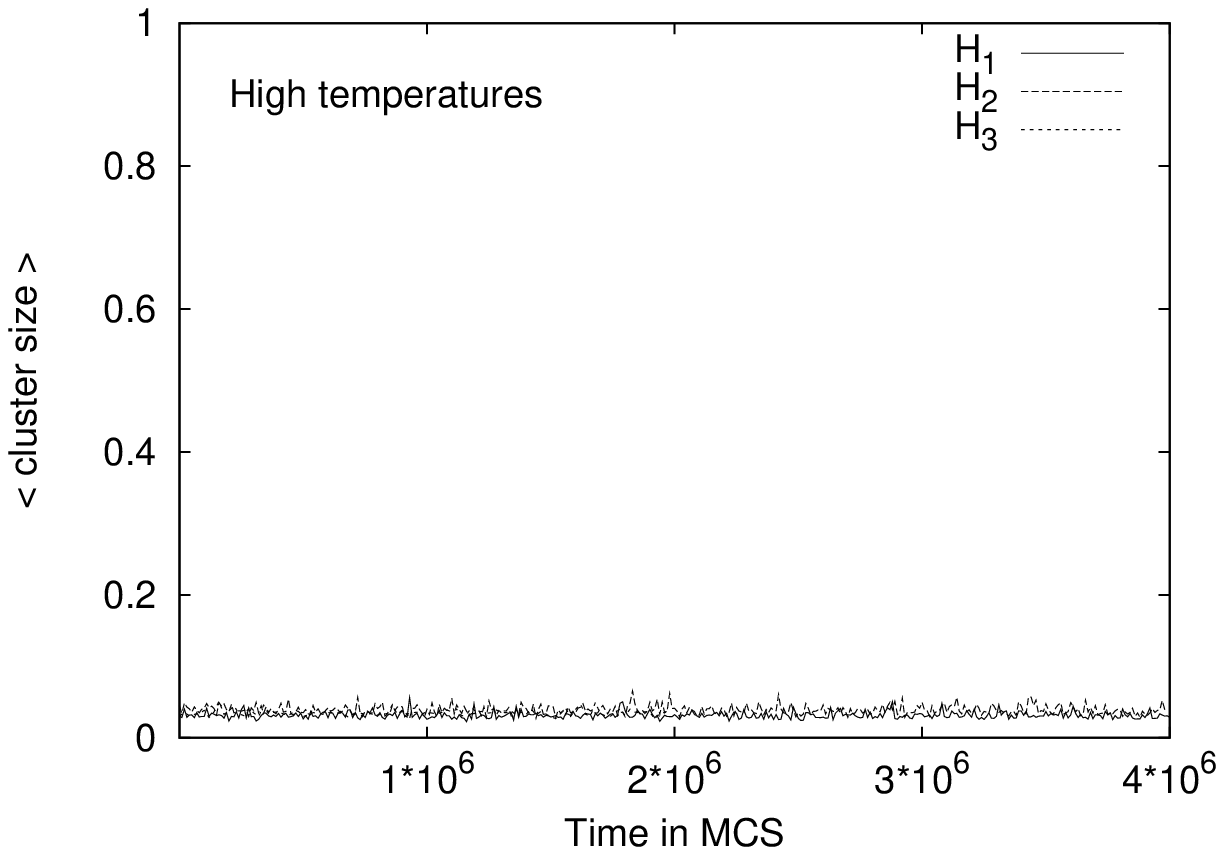}
\end{center}
\caption{Time dependence of the average number of clusters in the three models. Left upper panel - low temperatures, right upper panel - medium ones and bottom panel - high temperatures.}
\label{klastry}
\end{figure}

\begin{figure}
\begin{center}
\includegraphics[scale=0.5]{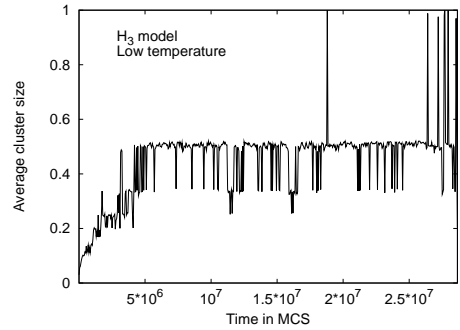}
\end{center}
\caption{Time dependence of the average cluster size for the $H_3$ model at low temperature.}
\label{Nklaster}
\end{figure}

As could be seen, the scaling is quite good, showing the same, or at least very similar, mechanisms, regardless of the range of interactions.  In the following we shall refer to temperatures $\tau \approx$ 0.80 as {\it low} temperatures, to $\tau \approx$ 1.02 as {\it medium} ones and when $\tau \approx$ 1.60 as {\it high} temperatures. In absolute values $T_c/J_1 \approx $ 0.44, 0.31 and 0.081 for the $H_1$, $H_2$ and $H_3$ models, respectively. From Fig.\ref{fi} we see that a better agreement is reached between the $H_1$ and $H_3$ (long range) models than between any of them and the $H_2$ model. This is supported by comparison of the spatial structures (snapshots) obtained at low temperatures after some $4 \times 10^6$ MCS (see Figure \ref{fig2})
.
We see that for $H_3$ and $H_1$ models a simple structure of the N\'{e}el type antiferromagnet is present, while for the $H_2$ model each particle has vacancies as its NN and particles of the opposite charge as its NNN..We see also that for short range interactions more particles are aggregated into a single cluster. Development of the spatial structure and its dynamics is shown in Fig.\ref{klastry}
At low temperatures the patterns for the $H_3$ and $H_1$ or $H_2$ models are different. In the $H_2$ model a single cluster is rather fast formed; it may break into two smaller ones, but they merge rather fast to form again one cluster. For the $H_1$ model formation of a single cluster takes a longer time, but then the behavior is similar to the $H_2$ model. We may therefore state that the asymptotic state for the two short-ranged models ($H_1$ and $H_2$) is a single cluster. Formation of a single cluster for the $H_3$ model, as seen from Fig.\ref{klastry}, is more difficult. We have run simulations in that case till $3 \times 10^7$ MCS, and as shows Fig.\ref{Nklaster} even after such long time the single cluster structure is nor very stable. The reason is that with short ranged interactions particles which do not belong to a cluster are basically free. They move randomly and when they come close to a cluster boundary they quickly join the cluster and stay there. Particles in the $H_3$ model are never really free and they experience, often contradictory, forces coming from many neighbors. Single particles in the $H_3$ model are rare since they quickly form pairs or quadruples, which move through the system at a slower rate. \\
At medium temperatures only interactions to NNN could lead to formation of a large clusters, which are nevertheless very unstable and quickly dissolve into smaller ones. In the case of the $H_3$ and $H_1$ models there is always a large number of rather small clusters and the respective curves lie very close one to another. Finally for high temperatures, well above $T_c$, as expected, there is no order whatsoever and no differences among the models.
\begin{figure}
\begin{center}
\includegraphics[scale=0.4]{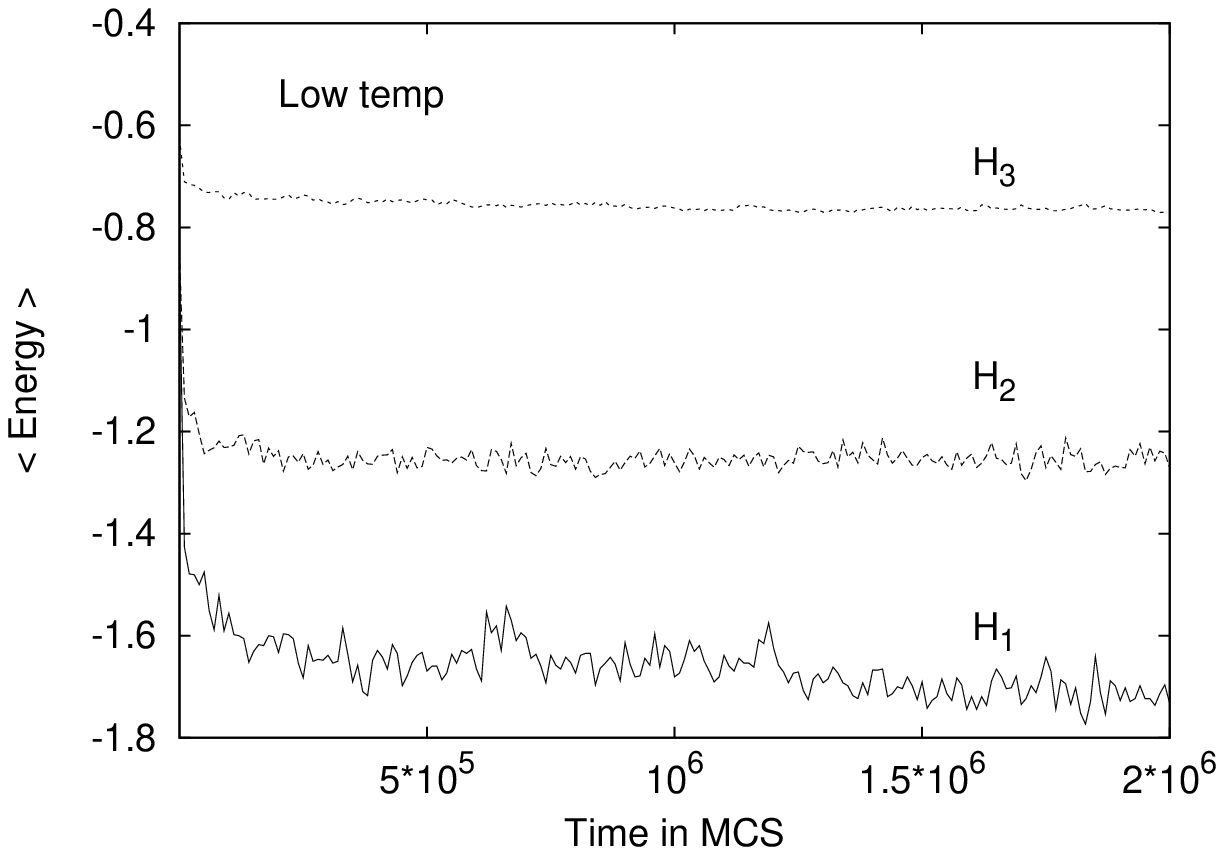}
\includegraphics[scale=0.4]{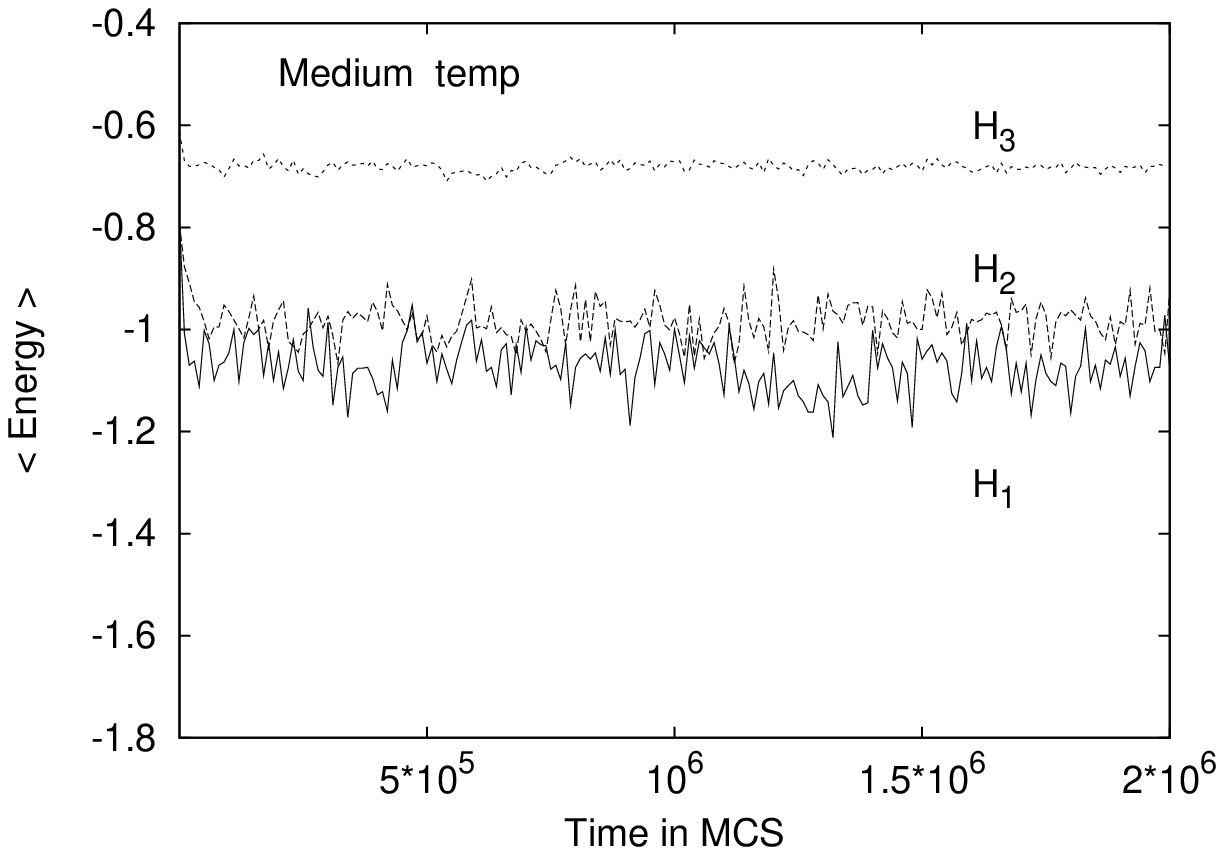}\\
\includegraphics[scale=0.4]{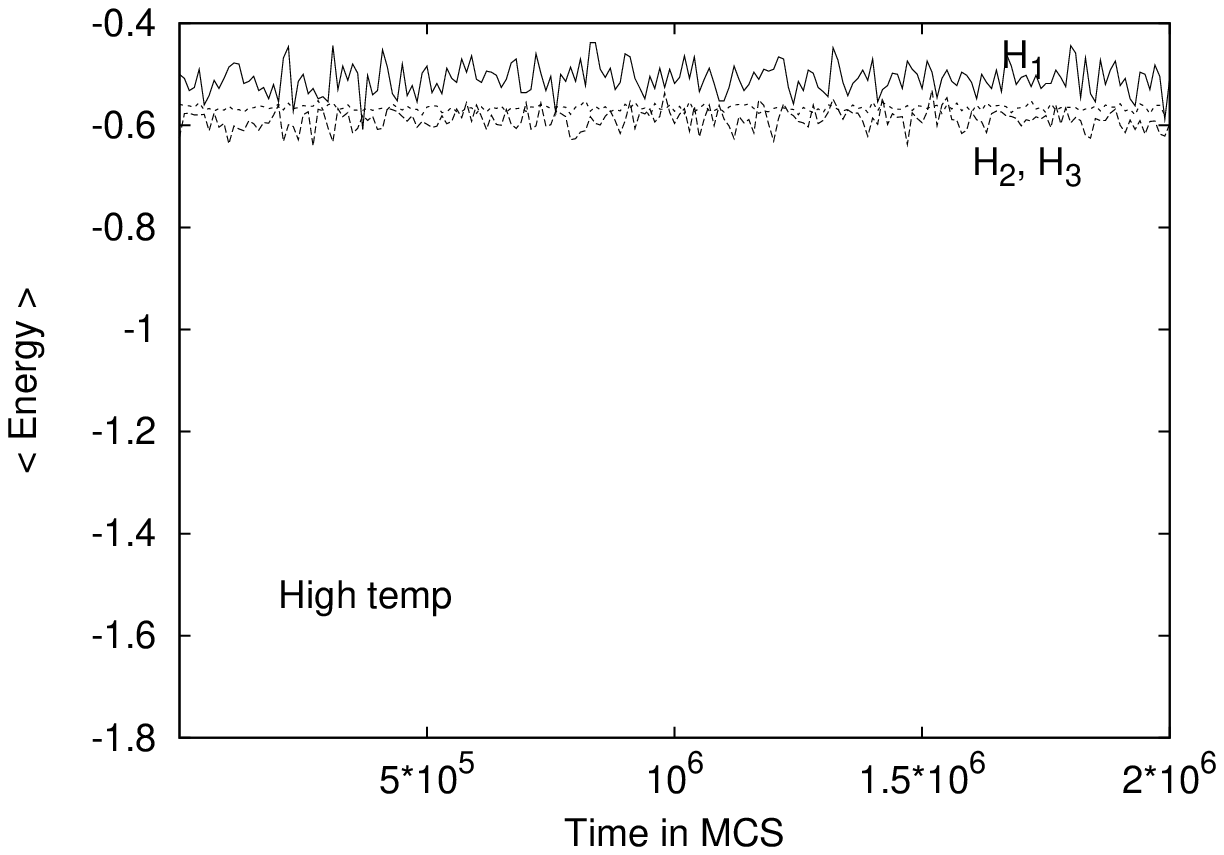}
\end{center}
\caption{Time dependence of the average energy in the three models. Left upper panel - low temperatures, right upper panel - medium ones and bottom panel - high temperatures.}
\label{energie}
\end{figure}
The dependence of the average energy on time is presented in Figures \ref{energie}. At low temperatures we see that the energies of all models are quite different. The lowest is the energy of the $H_1$ model, slightly above it lies the energy $H_2$, while the energy in the $H_3$ model is the highest. Most of the particles at such temperatures, belong to a rather well organized cluster, or clusters, with the lowest possible energy. Particles from the boundaries of the clusters do not have all partners, hence their contribution to the total (negative) energy is smaller. For the $H_1$ model, in general, only one or two partners are missing. For the $H_2$ model a particle interacts with more partners, hence the loss for the boundary particles is heavier and the total energy is raised more than for the $H_1$ model. The same, but to a larger extent, is true for the $H_3$ model. For higher (medium) temperatures all energies are shifted, the order is preserved, however the curves for $H_1$ and $H_2$ models are closer together than to the curve for the $H_3$ model. From Fig.\ref{klastry} we see that at medium temperatures large clusters are observed only for the $H_2$ model, hence the average energy in this case is less affected. Similarly for the $H_3$ model, where the change from the clusters size at low and medium temperatures is not dramatic. Raising the temperature has the largest effect on the $H_1$ model, where the large (single) cluster dissolved at higher temperatures into many smaller ones. There are many more particles on the borders of the clusters and the raise in the average energy is significant.  Finally, at temperatures well above $T_c$, all curves lie close together. As seen from Fig. \ref{klastry}  at those temperatures all larger clusters disappeared and only conglomerates of several particles remain.

\begin{figure}
\begin{center}
\includegraphics[scale=0.8]{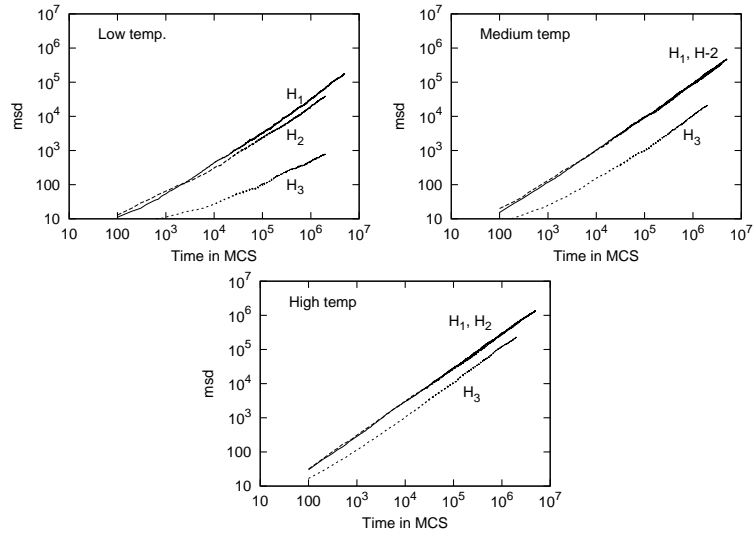}
\end{center}
\caption{Time dependence of the mean square displacement for the three models.}
\label{msd}
\end{figure}

Mean square displacement as a function of time is shown in Fig.\ref{msd} on a log-log scale for the three models and three sets of the reduced temperatures. It has always, at least after some initial period, a power type character:
\begin{equation}
\langle r^2 (t) \rangle \sim t^\alpha \,.
\end{equation}
At low temperatures $\alpha$ = 1.0 for the $H_1$ and $H_2$ models, hence diffusion has normal character, while for the $H_3$ model we have $\alpha$ = 0.60, indicating subdiffusion. The reason for this difference lies in contradictory forces acting on separate particles in the $H_3$ model, as well as in the fast formation of doublets and quadruplets. As we have shown before \cite{my1,my2} also in the $H_1$ and $H_2$ models subdiffusion is observed, but at temperatures lower than those reported here. For intermediate and high temperatures and after some $10^4$ MCS diffusion has normal character in each case.

\begin{figure}
\begin{center}
\includegraphics[scale=0.4]{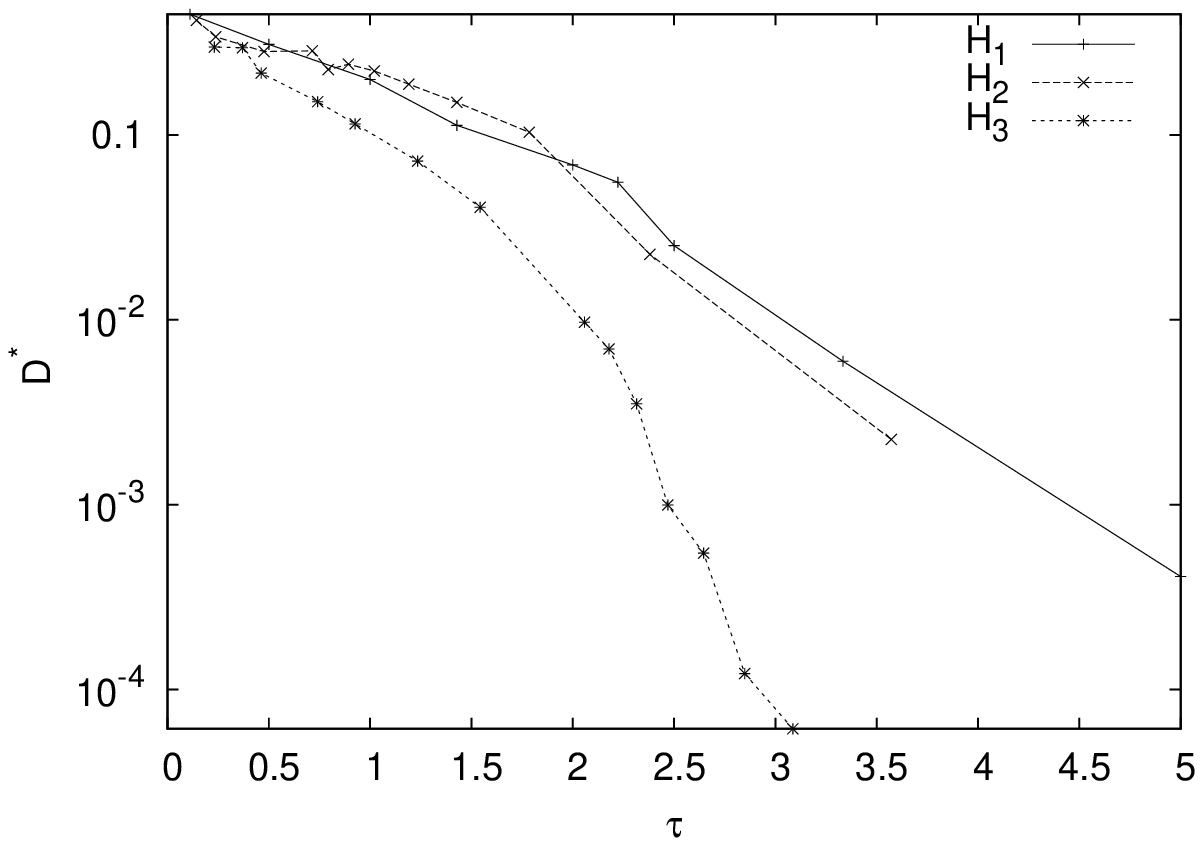}
\includegraphics[scale=0.4]{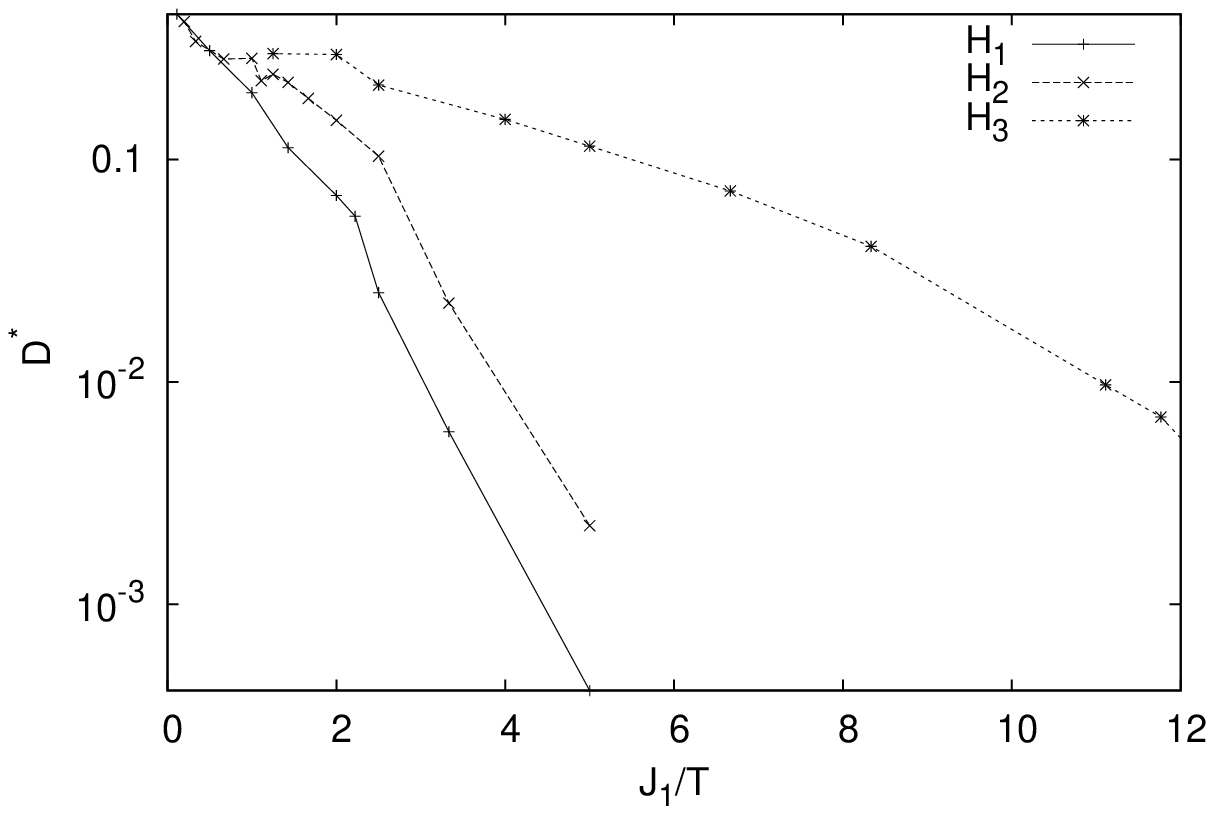}
\end{center}
\caption{Arrhenius plot of the tracer diffusion coefficient on scaled temperature (left panel) and unscaled (right panel) for the three models.}
\label{wspD}
\end{figure}

From the msd data the Arrhenius plots, i.e. the dependence of $D^*$ on $J_1/T$ could be constructed using Eq.(\ref{d}). In Fig.\ref{wspD} on left panel we present $D^*$ as a function of the reduced temperature $\tau$. As could be expected from the previous data and its analysis, diffusion in the $H_3$ model is slower, except at very high temperatures, where the particles could be treated as free. The dependence of $D^*$ on the temperature looks however different if we return to the unscaled temperature. Corresponding plot of $D^*$ versus temperature (divided by the interaction constant $J_1$) is shown in Fig.\ref{wspD} on right panel. Since the critical temperature for the $H_3$ model is much lower than the ones for the $H_1$ and $H_2$ models, at a given unscaled temperature, diffusion of particles subject to long ranged interactions is much faster.

\begin{figure}
\begin{center}
\includegraphics[scale=0.5]{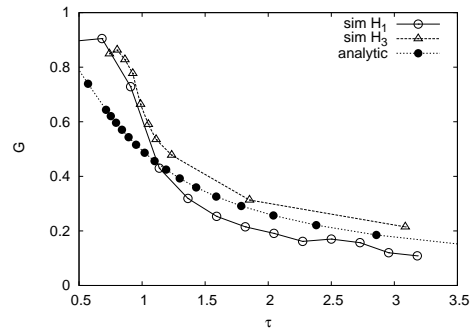}
\end{center}
\caption{Comparison of the two-point correlation function obtained for the $H_3$ and $H_1$ models via simulations and calculated analytically}
\label{porown}
\end{figure}
On the last plot (Fig.~(\ref{porown})) we present comparison of the two-point correlation function $G$, measuring correlations between particles which are NN, and defined as
\begin{equation}
\label{g1}
G \,=\, \frac{1}{4 N} \sum_{i=1}^N\sum_{j=1}^\delta n_i\,n_{i+j}  \, .
\end{equation}
The same correlation function could be calculated analytically, using the Oguchi method \cite{oguchi}, where the contribution from interactions of two NN particles is calculated exactly, while with all others is replaced by a mean field. Derivation of the formula for $G$ in the Oguchi method is described in \cite{my1}. Here we present only the final formulae
\begin{equation}
\label{g} G\,=\, 
\frac{\partial \ln Z}{\partial K}\,=\,  \frac{x^2 \sinh
K}{1+2x+x^2 \cosh K} \, .
\end{equation}
where $Z$ is the partition function and $x$ is
\begin{equation}
\label{x}
x\,=\, \frac{1 - 2c}{2(1-c)\cosh K}\left[\sqrt{1+ \frac{4c(1-c)\cosh K}{(1-2c)^2}} - 1 \right]  \, .
\end{equation}
 $K = J_1/kT$. $c$ is the coverage, which here equals 0.2. The curves presented  in Fig.\ref{porown} show similar character, except for low temperatures, where mean-field type theories are known to be not very reliable since they miss NN interactions which give significant contribution to correlations at low temperatures.
\section{Conclusions}
We have discussed the role of the range of interactions on static (spatial patterns, order parameter) and dynamic (msd, average cluster size, average energy) properties for the model introduced by us earlier \cite{my1,my2}. We consider here three versions of the model. In the first one ($H_1$) interactions are restricted to NN, in the second one ($H_2$) they are extended to NNN. In the third one ($H_3$), a particle interacts with all other particles not farther from it than half the linear size of the system, which with periodic boundary conditions used here means that one particle interacts practically with all other. The strength of the interactions falls off with the distance between the particles. We have shown that the static properties of the $H_3$ and $H_1$ models do not really differ -- both exhibit the same kind of ``antiferromagnetic'' ordering at low temperatures. Order parameter dependence on the temperature scaled by the respective critical one shows the same pattern for the three models. Similarity of the two point (to NN) correlation function behavior indicates that the long ranged interactions are, specially at low temperatures, screened.\\
Situation is however quantitatively different for dynamic features. Here taking a short (to NN or NNN) or long range range interactions yields different results. Organization of the system into one or two clusters takes much longer time for the $H_3$ model. Moreover such clusters are less stable, since there are more ``unsatisfied'' interactions for the border particles. Transition temperature from the ordered to disordered phase is much smaller for the $H_3$ than for the short range models, which also shows that the clusters formed in the $H_3$ model are less stable. Since $T_c$ for the $H_3$ model is much lower, at temperatures (unscaled) which are still below $T_c$ for the $H_1$ and $H_2$ models, particles with long range range interactions are nearly free. Hence their mobility is larger than those in the $H_1$ and $H_2$ models, still forming clusters.
At low (scaled) temperatures diffusion is slower in the case of the $H_3$ model, as practically on each particle act forces from many other particles and there are few, if any, free particles which could realize ordinary diffusive motion, characterized by simple $\langle r^2(t) \rangle \sim t $ dependence. Instead,  subdiffusion with an exponent equal 0.60 is observed. A system with long range range interactions has a tendency to organize itself, even at low (scaled) temperatures, in a set of smaller clusters than in the $H_1$ or $H_2$ models. Mobility of such groups of particles is lower than of separate particles encountered in the systems with short range interactions. Therefore. although finally, like in the other models, a single cluster appears, it takes, for the system of the size studied here, at least 10 times longer.\\
Our study does not necessarily mean that when modeling real systems of charged particles diffusing on a surface. always infinite range interactions should be taken. First of all, static properties are reasonably well described by short range models. Moreover, as we have demonstrated, the differences between the three versions of our model, have generally quantitative rather than qualitative character.  Therefore screening of the Coulomb like potential has, in the considered model, an important influence, reducing in some cases significantly the effective range of interactions. 

Acknowledgments\\
We are grateful to Z. Koza and M. Matyka for helpful discussions and comments.

\end{document}